\documentclass[11pt]{article}

\usepackage{graphicx}
\usepackage{color}

\begin{document}
\baselineskip 6.5mm

\title{The quantum Bernoulli map}
\author{Gonzalo Ordonez\footnote{Email: gordonez@butler.edu}\\
Physics and Astronomy Department,\\
 Butler University, 4600 Sunset Ave., Indianapolis, IN 46208 USA\\ 
 \\
Yingyue Boretz\\ 
Center for Complex Quantum Systems, \\
The University of Texas at Austin, Austin, TX 78712 USA }

\maketitle

\begin{abstract}
The classical Bernoulli and baker maps are two simple models of deterministic chaos. On the
level of ensembles, it has been shown that the time evolution  operator for these maps
admits generalized spectral representations in terms of decaying eigenfunctions. We introduce the
quantum version of the Bernoulli map. We define it as a projection of the quantum baker map. We construct
a quantum analogue of the generalized spectral representation, yielding quantum decaying states
represented by density matrices. The quantum decaying states develop a quasi-fractal shape limited
by the quantum uncertainty.  
\end{abstract}
%\pacs{05.45.Mt, 05.30.Ch, 05.70.Ln, 03.67.Lx}
\maketitle

%%%%%%%%%%%
\section{Introduction }
%%%%%%%%%%%%%

It is our privilege to contribute a paper to the memory of Prof. Shuichi Tasaki. On a few occasions Prof. Tasaki encouraged  us to publish the present work. We devote this work to him.

The present work concerns the subject of quantum chaos. More precisely, the study of quantum systems whose corresponding classical systems exhibit chaotic behavior. 
Prof. Tasaki was one of the first authors to link irreversibility and chaotic dynamics using  ``generalized spectral representations''  of time evolution operator (also called Frobenius-Perron or FP operator). One interesting outcome of the work of Prof. Tasaki and others (Refs.  \cite{Tasaki1}-\cite{Hir}, as well as \cite{Dean} and references therein) was the demonstration  that eigenfunctions of the FP operator may have a fractal nature. This was shown using classical chaotic maps, such as the multi-Bernoulli map or the multi-baker map. The present paper is motivated by this  work of Prof. Tasaki and others.

\begin{figure}
\begin{center}
\includegraphics[width=3in, angle=270]{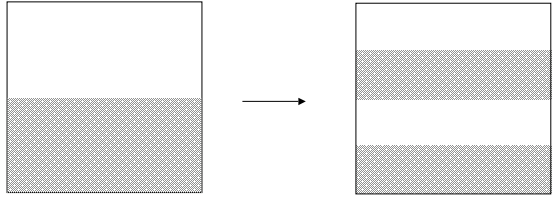}
\end{center}
\caption{The classical baker map.}
\label{fig1}
\end{figure}

In classical systems, chaos may appear in isolated systems with few degrees of freedom
as a consequence of stretching and folding dynamics. A simple model of this type of chaos
is the baker map. The baker map acts on a unit square with coordinates $(q, p)$ representing
the phase space. The square is squeezed down in $p$ direction; it is stretched in $q$ direction by
a factor of $2$ and then right half is put on top (see Fig. \ref{fig1}). These time-evolution rules are
a simple example of stretching and folding dynamics. Their consequence is a chaotic time
evolution where any uncertainty in the initial condition grows exponentially with time until
the uncertainty is spread over the whole phase space.

Despite its chaotic evolution, the baker map is invertible and unitary. After applying the
map any number of times, the resulting final phase-space distribution can be reverted to the
initial distribution by application of the inverse map.
%%%
%%%

An even simpler map is obtained by projecting the baker map onto the horizontal $(q)$
axis. This simpler map is called the Bernoulli map; in contrast to the baker map it is not
invertible. The Bernoulli map maps a number $q_\tau \in 
(0,1]$ as
\begin{equation}
  q_{\tau+1} =2 q_\tau \,\, {\rm mod}\, 1.
\end{equation}
where $\tau$ is a discretized time variable, taking only integer values.

One can understand the chaotic nature of the Bernoulli map by expressing the variable $q$ in
binary notation. Written this way, the effect of the Bernoulli map is simply a digit shift. If
$q_\tau = 0.d_1 d_2 d_3 ...$ then $q_{\tau+1} = 0.d_2 d_3 d_4 ....$, where $d_i = 0$ or $1$. An initial $q$ given by an irrational
number with random digits will be mapped to a random new point; the trajectory followed
by $q$ over time will be a random trajectory.

Corresponding to this behavior of a single trajectory, an ensemble of trajectories initially
close together will end up uniformly spread out over the whole phase space (i.e., over the
whole unit interval). An ensemble uniformly distributed over the whole phase space is an
equilibrium distribution, since further applications of the map will not change it. Equilibrium
is an attractor for ensembles; any initially smooth distribution of trajectories will approach
the equilibrium distribution. This situation is reminiscent of the irreversible approach to
equilibrium of ensembles in macroscopic systems, exhibited in processes such as diffusion.

In fact, as discussed in Refs.  \cite{Tasaki1,Hir,Dean} irreversibility (or time-symmetry breaking) can
be formulated explicitly in simple chaotic models, such as the Bernoulli and baker maps.
In precise terms, irreversibility means that the FP operator for ensembles  admits eigenvalues with  magnitude less than $1$, which implies that the corresponding eigenfunctions decay in time.
Such eigenvalues appear within ``generalized spectral representations'' of the FP operator. A remarkable feature of these representations is that they may include decaying eigenfunctions even in cases where the FP operator is unitary, as is the case in the classical baker
map. These representations involve both regular functions and generalized functions (or functionals) playing the role of right and left eigenstates of the FP operator, respectively.

In quantum mechanics, chaos is a more involved concept. For small, isolated quantum
systems the Hamiltonian has a discrete real spectrum. The eigenvalues of the FP operator
are complex numbers with magnitude equal to $1$; no approach to equilibrium is apparent
because there are no decaying eigenfunctions.

To describe quantum chaos one can introduce certain criteria, such as integrability, or the
spacing of energy levels \cite{Reichl} (a thorough discussion on integrability of quantum systems is given in Ref. \cite{Cirelli}).  In this paper, we will follow a different approach. We will  consider quantum maps, where the time evolution is applied in steps  (see Refs. \cite{Sud1,Rau,Sud2} for a discussion on quantum stochastic maps). We
will study density matrices, corresponding to the classical statistical ensembles. And we
will assume the quantum maps are coupled to their environment, which causes instantaneous
loss of coherence at each time step \cite{Paz,Valeri}. 

Specifically, we will study the quantum baker map coupled to its environment. The
quantum baker map is the quantum analogue of the classical baker map. It has a physical
realization as a set of quantum bits that can be shifted by an external action  \cite{Brun}, one step at a time.  We will assume that the environment causes the quantum bits to experience instant decoherence after each application of the quantum map. The result of the combination of the quantum baker map with decoherence will be a new map, which will turn out to be the quantum analogue of the classical Bernoulli map. We
will call this new map the quantum Bernoulli map.

We will show that in the quantum Bernoulli map density matrices approach equilibrium, similar to their classical counterparts. 
The approach to equilibrium will be described by introducing quasi-eigenstates of the quantum FP operator, which decay in time.  These states will appear in an expansion of the quantum FP operator analogous to the generalized spectral representation of the classical FP operator. 

The present work is of interest because it exemplifies the quantum correspondence between classical chaotic maps and their quantum analogues using an approach that as far as we know has not been taken before, namely through the generalized spectral representation of the FP operator. We will show that the quasi-eigenstates of the quantum FP operator approach their classical counterparts in the  classical limit. Interestingly, when approaching this limit, the eigenfunctions take a self-similar, quasi-fractal shape up to the point allowed by Heisenberg's uncertainty principle. (Fractals and self-similarity in quantum systems have been studied in Refs. \cite{Reichl2}-\cite{Berry}, although in different contexts. A connection
between the decaying eigenstates of the classical FP operator and spectral properties of
chaotic quantum systems has been investigated in Ref. \cite{And}).

In the following section we will  summarize the ensemble description of the classical
Bernoulli map, including the formulation of decaying eigenstates of the FP operator.
Then we will describe the quantum Bernoulli map and we will formulate the quantum analogue of the generalized spectral representation of the FP operator. We will finish by discussing the development of the quasi-fractal shape in the quasi-eigenstates and giving final remarks.

%%%%%%
\section{The classical  Bernoulli map}
\label{CB}
%%%%%%%

In this section we will review the classical Bernoulli map, following Ref. \cite{Dean}. We will focus
on an ensemble description of the maps, where the central role is played by the
FP operator and its spectral (eigenvalue) representation. As we will see the  right eigenstates of the FP operator are smooth functions that decay in time, while  the left-eigenstates are functionals. The set of right and left eigenstates together with the corresponding eigenvalues will form the generalized spectral
representation of the FP operator.

To introduce the Bernoulli map we start with the baker map illustrated in Fig. \ref{fig1}. Its FP
operator is defined as
\begin{equation}
U_b \rho(q,p) =  \left\{ \begin{array}{ll}
       \rho(q/2, 2p), &  0\le p < 1/2\\
        \rho(q/2+1/2, 2p-1),  &  1/2\le p <1.
                 \end{array} \right. \nonumber\\
\end{equation}
for a probability density $\rho(q,p)$ defined in the unit square.   The Bernoulli map is a projection  of the baker map, obtained  by integrating over a re-scaled $p$,
\begin{equation}
\int_0^1 dp\rho (q,p)=\frac{1}{2}\int_0^{1} dp\,
\rho \left( \frac{q}{2} ,p \right) + \frac{1}{2} \int_{0}^1  dp\, \rho\left(\frac{q+1}{2},p\right)
\end{equation}
This gives the FP operator for the Bernoulli map as 
\begin{equation}
U_B \rho (q)=\frac{1}{2}\left[\rho \left(\frac{q}{2}\right)+\rho \left(\frac{1+q}{2}\right)\right] 
\end{equation}
where $\rho(q)$ is a probability density defined in the unit interval. Successive applications of $U_B$
on $\rho(q)$ will make it evolve towards  
the uniform equilibrium distribution $\rho^{\rm eq}(q) = 1$.
This is true provided $\rho(q)$ is a normalizable function, which excludes delta-functions corresponding
to trajectories. To define normalizable functions we will use the inner product
\begin{equation}
\label{inner}
\left( X |Y\right)=  \int_0^1 dq X^*(q) Y(q)
\end{equation}
where $Y(q) = \left( q |Y\right)$ and $X^*(q) = \left( X|q\right)$. The norm of $\rho$ is then $\left(\rho |\rho\right)$.

We are discussing the approach to equilibrium of probability densities. But there is  a property of the Bernoulli map that seems to indicate that there can be no approach
to equilibrium at all! Solving this apparent contradiction will lead us to the main result of  this section. We first introduce the Hermitian conjugate operator $U_B^\dagger$
in the usual way
\begin{equation}
\label{inner2}
\left( X |U_B^\dagger| Y\right) = \left( Y |U_B|X\right)^*
\end{equation}
This equation shows that the right eigenstates of $U_B^\dagger$ must be left eigenstates of $U_B$ and vice-versa. Using the explicit form of $U_B$ the explicit form of $U_B^\dagger$ can be shown to be
\begin{equation}
 U_B^\dagger Y(q)=
 \left\{ \begin{array}{ll}
        Y(2q), &  0\le q < 1/2\\
        Y(2q-1),  & 1/2\le q <1.
                 \end{array} \right. \nonumber\\
\end{equation}
The  property of the   Bernoulli map that concerns us here is that while $U_B^\dagger$ is not unitary because $U_B^\dagger U_B \ne 1$,  it is isometric:
\begin{equation}
\label{UBUB}
U_B U_B^\dagger = 1
\end{equation}
This is an important property because it means that any right eigenfunctions of $U_B^\dagger$ that are normalizable must have eigenvalues with magnitude equal to one.  
To see this, note that $\left( \rho| U_B U_B^\dagger |\rho\right) = \left(\rho|\rho\right) =1$ for any $\rho$ with unit norm, including any normalizable eigenstates of $U_B^\dagger$.

The isometry of $U_B^\dagger$ implies that it has  no normalizable, decaying right-eigenfunctions, because normalizable eigenfunctions must have eigenvalues with unit magnitude. This, in turn, implies that there are no normalizable, decaying, left eigenfunctions of $U_B$. All this seems to negate the approach to equilibrium characterized by decaying eigenfunctions. However, there is a way out. As shown in Refs. \cite{Tasaki1, Hir,Dean}, $U_B$ can have decaying left eigen-{\it functionals}. In fact, the functionals ${\tilde {B}_\alpha}$ (with $\alpha$ a non-negative number) defined by
\begin{equation}
\label{lefte}
\left( {\tilde {B}_\alpha} |f \right)
= \frac{1}{\alpha!}\int_0^1 dq\,\frac{d^\alpha}{dq^\alpha} f(q)
\end{equation}
are left eigenstates of $U_B$ with eigenvalues $2^{-\alpha}$, which means that 
\begin{equation}
\left( {\tilde {B}_\alpha} 
\right|U_B = \frac{1}{2^\alpha} \left({\tilde {B}_\alpha } \right|.
\end{equation}
Taking the Hermitian conjugate we obtain
\begin{equation}
U_B^\dagger \left. | {\tilde B_\alpha} \right)
= \left. \frac{1}{2^\alpha}| {\tilde {B}_\alpha} \right),
\end{equation}
which means that $\left.| {\tilde B_\alpha} \right)$ are right eigenstates of $U_B^\dagger$ with eigenvalue $2^{-\alpha}$. Thus allowing functionals to play the role of eigenfunctions allows for the existence of eigenvalues smaller than 1. 

The operator $U_B$ also has eigenvalues with magnitude less than $1$. However, the corresponding right eigenfunctions {\it are} normalizable. Indeed, since $U_B^\dagger U_B \ne 1$, the argument below Eq. (\ref{UBUB}) does not apply to $U_B$. As a result, $U_B$ can have normalizable right eigenfunctions with eigenvalues with magnitude under 1.  We will discuss these eigenfunctions now.

The right eigenfunctions of $U_B$ can be constructed by noting that $U_B$ acts as a shift operator
\begin{equation}
\label{UBe1}
 U_B  e_{j,l}(q) =
 \left\{ \begin{array}{ll}
        e_{j-1,l}(q), &  j\ge 1\\
        0,  &  j=0,
                 \end{array} \right. \nonumber\\
\end{equation}
and
\begin{equation}
\label{UBe}
 U_B^\tau  e_{j,l}(q) =
 \left\{ \begin{array}{ll}
        e_{j-\tau,l}(q), &  j\ge \tau\\
        0,  &  j< \tau.
                 \end{array} \right. \nonumber\\
\end{equation}
on the Fourier-basis functions
\begin{equation}
\label{edef}
e_{j,l}(q) = \exp [2\pi i 2^j(2l+1) q]
\end{equation}
with $j\ge 0$ and  $-\infty <l<\infty $ integers. 

The existence of shift states allows the construction of a type of coherent eigenstates of $U_B$, given by
\begin{equation}
\label{Bdef}
B_\alpha(q) \equiv - \sum_{j,l}\frac{\alpha!}{[2\pi i 2^j(2l+1)]^\alpha} e_{j,l}(q)
\end{equation}
It turns out these are the  Bernoulli  polynomials $B_\alpha(q)$ of degree $\alpha$ \cite{Tasaki1,Hir,Dean}. We have 
\begin{equation}\label{UBB}
U_B \left| 
{B_\alpha } \right) = \frac{1}{2^\alpha}  \left| 
{B_\alpha } \right) 
\end{equation}

Summarizing, while the right eigenstates of $U_B$ are smooth, normalizable functions (polynomials), the right eigenstates of $U_B^\dagger$ (and thus the left eigenstates of $U_B$) are non-normalizable functionals. In order to obtain eigenvalues of $U_B^\dagger$ with magnitude less than $1$ we had to leave the domain of regular functions. This is an interesting point. It ties irreversibility to a mathematical formulation involving extended function spaces. 

The Bernoulli polynomials and the functionals $ \left({\tilde B}_\alpha\right|$ are orthogonal,
\begin{equation}
\left( {\tilde {B}_\beta} | B_\alpha\right) = \delta_{\beta,\alpha}.
\end{equation}
As a result, in the domain where $\sum_\alpha  \left| B_\alpha \right) \left({\tilde B}_\alpha\right| =1$, the FP operator can be represented as 
\begin{equation}
\label{grep}
U_B^\tau \rho(q) =  \sum_{\alpha=0}^\infty 2^{-\alpha \tau} B_\alpha(q) \left( {\tilde {B}_\alpha} | \rho \right)
\end{equation}
where $B_0(q) = 1$.  This is the generalized spectral representation of the FP operator. The representation (\ref{grep}) displays the decay rates $2^{-\alpha}$, which are powers of the Lyapounov exponent  $\exp(\ln 2)$ of the Bernoulli map.  This representation clearly shows that when $\tau\to \infty$ probability densities  approach the uniform, equilibrium density $\rho^{\rm eq} = B_0(q) =1$. As discussed in \cite{Tasaki1, Hir,Dean} the representation (\ref{grep}) is valid for smooth, differentiable probability densities (which can be expressed as superposition of Bernoulli polynomials). As already mentioned, this excludes trajectories, which are represented by Dirac delta functions. 

Setting $\tau=0$ in Eq. (\ref{grep}) and evaluating the inner product  $ \left( {\tilde {B}_\alpha} | \rho \right)$ explicitly we obtain, with $ \rho^{(\alpha)}(q)  \equiv d^\alpha \rho/dq^\alpha$, and $\rho_0 \equiv \left( {\tilde {B}_0} | \rho \right)$
\begin{eqnarray}
\label{grepq0}
\rho(q) &=& \rho_0 + \sum_{\alpha=1}^\infty \frac{1}{\alpha!}  B_\alpha(q) \left[\rho^{(\alpha-1)}(1) -   \rho^{(\alpha-1)}(0)\right] 
\end{eqnarray}
This is another form of the generalized spectral representation (\ref{grep}) for $\tau=0$. It is also known as the Euler-Maclaurin  expansion.  It is the generalized spectral representation of the unit operator $U_B^\tau=1$ for $\tau=0$. For later comparison with the quantum Bernoulli  map we write this expansion as 
\begin{eqnarray}
\label{grepq}
\rho(q) &=& \rho_0 \\
&+&  \sum_{\alpha=1}^\infty \frac{1}{\alpha!} \left[ B_\alpha(q) \rho^{(\alpha-1)}(1) -  (-1)^\alpha B_\alpha(1-q) \rho^{(\alpha-1)}(0)\right] \nonumber
\end{eqnarray}
where we used the symmetry property of the Bernoulli polynomials,
\begin{equation}
 B_\alpha(q) = (-1)^\alpha B_\alpha(1-q).
 \end{equation}
For $\tau>0$ the generalized spectral representation (\ref{grep}) takes the form
\begin{eqnarray}
\label{grepqt}
U_B^\tau \rho(q) &=& \rho_0 \\
&+&  \sum_{\alpha=1}^\infty \frac{1}{2^{\alpha \tau}} \frac{1}{\alpha!} \left[ B_\alpha(q) \rho^{(\alpha-1)}(1) -  (-1)^\alpha B_\alpha(1-q) \rho^{(\alpha-1)}(0)\right] \nonumber
\end{eqnarray}
%%

%%%%%%%%%%%%%%%%
\section{Quantum baker map}
\label{QBM}
%%%%%%%%%%%%%%%%%%

In the following we will   focus our attention on the quantum version of the Bernoulli map and the representation corresponding to Eqs. (\ref{grep}) or (\ref{grepqt}).  Before going to the quantum Bernoulli map, in this section we  will give a brief introduction to the quantum baker map, upon which the quantum Bernoulli map will be built.

We will follow the definition of 
the quantum baker map given  in Ref.  \cite{Bal}. As in the classical case, we have a unit square, with horizontal coordinate $q_n$ (``position'') and vertical coordinate $p_m$ (``momentum''). Since this is a closed, finite quantum system, these coordinates are quantized. We divide the 
unit square into an $N\times N$ grid. We assume that $N = 2^D$ where $D$ is an integer. The quantized position and momentum are given by
\begin{equation}
q_n =\frac{n}{N},\quad n = 0, 1, 2 {\ldots}N-1
\end{equation}
\begin{equation}
p_m =\frac{m}{N}, \quad m = 0, 1, 2 {\ldots}N-1
\end{equation}
The position states $|q_n\rangle$ or
the momentum states  $|p_m\rangle$ are a basis for all possible states of the system. They satisfy the orthogonality  relations
\begin{equation}
\left\langle {q_n } \mathrel{\left| {\vphantom {{q_n } {q_{{n}'} }}} \right. 
\kern-\nulldelimiterspace} {q_{{n}'} } \right\rangle 
= 
\delta_{n,n'}, \qquad \left\langle {p_m } \mathrel{\left| {\vphantom {{p_m } {p_{{m}'} }}} \right. 
\kern-\nulldelimiterspace} {p_{{m}'} } \right\rangle 
= \delta_{m,m'}
\end{equation}
and completeness relations
\begin{equation}
\label{eq4}
\sum\limits_{n=0}^{N-1} {\left| {q_{{n}} } \right\rangle } \left\langle 
{q_n } \right|=1
\end{equation}
\begin{equation}
\label{eq3}
\sum\limits_{m=0}^{N-1} {\left| {p_{{m}} } \right\rangle } \left\langle 
{p_m } \right|=1
\end{equation}
The transformation from the position to momentum representation is given by 
\begin{eqnarray}
\left\langle {q_n } \mathrel{\left| {\vphantom {{q_n } {p_m }}} \right. 
\kern-\nulldelimiterspace} {p_m } \right\rangle 
 &=& \frac{1}{\sqrt N }\exp[(ip_m q_n )/\hbar ]  \nonumber\\
 &=& \frac{1}{\sqrt N } \exp[ (2\pi imn)/N]
\end{eqnarray}
where 
\begin{equation}
 \hbar =\frac{1}{2\pi N}
\end{equation}
plays the role of Planck's constant in this model. The classical limit is given by $N\to\infty$, corresponding to an infinitely fine grid.

The quantum baker map is represented by a unitary operator $T$. In position 
representation its matrix elements are
\begin{equation}\label{Tdef}
\left\langle {q_{{n}'} } \mathrel{\left| {\vphantom {{q_{{n}'} } {\left. T 
\right|q_n }}} \right. \kern-\nulldelimiterspace} {\left. T \right|q_n }  \right\rangle 
=
{\frac{\sqrt 2 }{N}} \times
 \left\{ \begin{array}{ll}
       \sum_{m=0}^{N/2-1} 
 \exp \left[ {2\pi im({n}'-2n)/N} \right] , &  0\le n\le  N/2\\
       \\
        \sum_{m=N/2}^{N-1} 
 \exp \left[ {2\pi im({n}'-2n)/N} \right],  &  N/2 \le n<N.
                 \end{array} \right. \nonumber\\
\end{equation}
States $|\psi_\tau\rangle$ are transformed as 
\begin{equation}
|\psi_{\tau+1}\rangle =  T |\psi_\tau\rangle
\end{equation}
Hence density operators $\rho_\tau$ are transformed as
\begin{equation}
\rho_{\tau+1}  =  T \rho_\tau T^\dagger
\end{equation}
This defines the FP operator for the quantum baker map. In the classical limit the quantum baker map goes to the classical baker map \cite{Bal}.

%%%%%%%%%%
\section{Quantum Bernoulli map}
\label{QBiM}
%%%%%%%%%%%%%%
We will assume that the quantum baker map experiences  instantaneous decoherence in $|q_n\rangle$ representation.   As a result, the off-diagonal 
terms of the density matrix will disappear instantaneously after each time step. Therefore, we will only keep the 
diagonal terms at each step of time evolution. The two-dimensional quantum 
baker map will be projected into a one-dimensional map, which we will call quantum Bernoulli map.

The FP operator  $U_B^Q $ for  the quantum Bernoulli map is thus defined as
\begin{eqnarray}\label{UQdef}
\rho_{\tau+1} &=& U_B^Q \rho_\tau = \left[T \rho_\tau T^\dagger\right]_{\rm diagonal} \nonumber\\
&=&
\sum\limits_{n=0}^{N-1} {\sum\limits_{{n}'=0}^{N-1} {\left| {q_{{n}'} } 
\right\rangle } } \left\langle {q_{{n}'} } \right|T\left| {q_n } 
\right\rangle \left\langle q_n \left| \rho_\tau\right|q_n\right\rangle \left\langle {q_n } \right|T^\dagger\left| 
{q_{{n}'} } \right\rangle \left\langle {q_{{n}'} } \right| 
\end{eqnarray}
The quantum Bernoulli map acts on the diagonal component of density matrices, and transforms them into new diagonal density matrices.

To see the quantum-classical correspondence of the Bernoulli map we start by defining shift states as 
\begin{equation}
{ e}_{j,l}= \sum_{n=0}^{N-1} |q_n\rangle e_{j,l}(q_n) \langle q_n|
\end{equation}
where $e_{j,l}(x)$ is defined in Eq. (\ref{edef}).  Due to the discrete nature of the coordinate $q_n$, the functions
\begin{equation}
e_k(q_n) \equiv  e_{j,l}(q_n), \quad k = 2^j(2l+1)
\end{equation}
satisfy the relation
\begin{equation}
e_k(q_n) = e_{k\pm N} (q_n).
\end{equation}
Hence from now on we restrict the domain of $k$ (and thus of $j$ and $l$) in such a way that 
\begin{equation}
\label{kdom}
 -N/2 \le k < N/2.
\end{equation}
This domain goes to the classical domain in the limit $N\to\infty$. 

Note that because the number $N$ of modes $k$ is finite, the number of shift states is finite. As a result, the quantum Bernoulli map is not isometric. The limited number of shift states prevents the construction of exact eigenstates of the FP operator as was done for the classical Bernoulli map. However, we will find quasi-eigenstates of the quantum Bernoulli map, which will approach exact eigenstates in the classical limit.

For $k=0$, the shift state $e_0(q_n) = 1$ is just a constant and we have
\begin{equation}
\label{conste}
(U_B^Q )^\tau { e}_{0} (q_{n} ) = { e}_{0} (q_{n} ).
\end{equation}
For $k\ne 0$ we use Eqs. (\ref{Tdef}) and  (\ref{UQdef})  to obtain \cite{Yingyue}
\begin{equation}
\label{UBQee}
(U_B^Q )^\tau { e}_{j,l} (q_{n} ) = 
 \left\{ \begin{array}{ll}
        { e}_{j-\tau,l} (q_{n} ) \prod_{\tau'=0}^{\tau-1} s_{j-\tau',l}(n,N) , &  j\ge \tau\\
        0,  &  j<\tau.
                 \end{array} \right. \nonumber\\
\end{equation}
where
\begin{equation}
\label{weight_s}
 s_{j',l}(n,N) = 
 \left\{ \begin{array}{ll}
        1 , & {\rm for\,} n \, {\rm even}\\
        1 - 2^{j'+1} |2l +1|/N,  &  {\rm for\,} n \, {\rm odd}.
                 \end{array} \right. \nonumber\\
\end{equation}
The result (\ref{UBQee}) says that FP operator acts as a weighted shift on $e_{j,l}(q_n)$. To obtain a non-weighted shift we define the new shift states
\begin{equation}
\label{etild}
 \tilde {e}_{j,l} (q_n )=\mu _{j,l} (n,N)e_{j,l} (q_n ), 
 \end{equation}
 where
\begin{equation}
\mu _{j,l} (n,N) =  \left[\prod_{j'=1}^j s_{j',l}(n,N)\right]^{-1}
\end{equation}
The new shift states satisfy
\begin{equation}
\label{UBQe}
(U_B^Q )^\tau {\tilde e}_{j,l} (q_{n} ) = 
 \left\{ \begin{array}{ll}
        \tilde {e}_{j-\tau,l} (q_{n} ), &  j\ge \tau\\
        0,  &  j<\tau.
                 \end{array} \right. \nonumber\\
\end{equation}
Up to this point we have a complete quantum-classical correspondence (compare Eqs. (\ref{UBe}) and (\ref{UBQe})). The weight factor in Eq. (\ref{weight_s}) has the form
\begin{equation}\label{shbar}
 s_{j,l}(n,N) = 1 + O(\hbar)
\end{equation}
so it goes to $1$ in the classical limit (i.e., $N\to \infty$ with $l, j$ finite) and we recover Eq. (\ref{UBe}).

%%%%%%%%%%%%%%%
\section{The quantum analogue of the generalized spectral representation for $\tau=0$}
\label{EM}
%%%%%%%%%%%%%%%%%

We will study now the quantum-classical correspondence of the Bernoulli map based on the generalized spectral representation discussed in Section \ref{CB}. We will first find an expansion of the unit operator $(U_B^Q)^\tau$ with $\tau=0$,  analogous to the Euler-Maclaurin expansion in Eq.   (\ref{grepq}). In the next section we will consider the case $\tau>0$.

Consider an arbitrary density $\rho(q_n)$.  The uniform part is given by
\begin{equation}
\rho_0 = \frac{1}{N}\sum\limits_{n' =0}^{N-1} \rho (q_{n'})
 \end{equation}
The non-uniform part,
\begin{equation}
\label{nonupart}
 \delta\rho(q_n) = \rho(q_n) - \rho_0 \end{equation}
is  expanded   in terms of the shift states $e_{j,l} (q_{{n}'} )$ as
\begin{eqnarray} \label{QGR1}
\delta\rho (q_n) = \frac{1}{N} 
\sum_{{n}'=0}^{N-1} \sum_{j,l}^{[N]}
 e_{j,l}  (q_n) e_{j,l}^*(q_{n'})\rho (q_{n'})
\end{eqnarray}
where the superscript $[N]$ indicates the restriction in Eq. (\ref{kdom}).  In the classical case the generalized spectral  representation involves derivatives. Seeking a correspondence, in the quantum case we introduce differences. We use the notations
\begin{equation}
  {\rho }^{(\alpha)} (q_{n'},N)= N \left[\rho^{(\alpha-1)}(q_{n'},N)-\rho^{(\alpha-1)}(q_{n'-1},N)\right],
  \quad \alpha>0
\end{equation}
and 
\begin{equation}
  {\rho }^{(0)} (q_{n'},N)= {\rho} (q_{n'})
\end{equation}
for the  differences. In the classical limit, with the condition $\alpha\ll N$,  the differences  go to derivatives,
\begin{equation}\label{dev2diff}
 \lim_{N\to\infty}  {\rho }^{(\alpha)} (q,N)=  \frac{d^\alpha}{dq^\alpha} {\rho} (q)
\end{equation}
Note that the difference at a point $q_{n'}$ gives the function at that point minus the function at the point $q_{n'-1}$ on the left. We could as well have taken the difference between the points $q_{n'+1}$ and $q_{n'}$. As we will see, our definition of difference breaks the symmetry of the eigenfunctions around $1/2$ (the mid point between the end  points $0$ and $1$) that exists in the classical model. The symmetry is restored in the classical limit. 

The summation over $n'$ in Eq. (\ref{QGR1}) may be written as
\begin{eqnarray}
\label{EM1}
  R_0 \equiv \sum_{{n}'=0}^{N-1} 
  e_{j,l}^*(q_{n'})\rho (q_{n'})  
=  \frac{-1}{1-e_{j,l}^*(q_1)} \left\{  \rho (q_{N-1})-\rho (q_0)\right\}  + R_1
  \end{eqnarray}
where
\begin{equation}
\label{R1}
R_1 =  \frac{N^{-1}}{1-e_{j,l}^*(q_1)} 
\sum\limits_{{n}'=1}^{N-1} e_{j,l}^*(q_{n'})\rho^{(1)} (q_{n'},N)
\end{equation}
The summation in $R_1$ is similar to the left hand side of Eq.  (\ref{EM1}). Thus we have
\begin{eqnarray}
\label{EM2}
  R_1 
=  \frac{-N^{-1}}{\left(1-e_{j,l}^*(q_1)\right)^2} 
  \left\{   \rho^{(1)} (q_{N-1},N)-\rho^{(1)}(q_1,N)e_{j,l}^*(q_1)\right\}  + R_2
  \end{eqnarray}
where
\begin{equation}
\label{R2}
R_2 =  \frac{N^{-2}}{\left(1-e_{j,l}^*(q_1)\right)^2}
\sum\limits_{{n}'=2}^{N-1} e_{j,l}^*(q_{n'})\rho^{(2)} (q_{n'},N)
\end{equation}
By recursion we get 
\begin{eqnarray}
&& \sum_{n'=0}^{N-1} e_{j,l}^*(q_{n'})\rho (q_{n'}) 
= \sum\limits_{\alpha =1}^N  \frac{-N^{-(\alpha-1)}}{\left(1-e_{j,l}^*(q_1)\right)^\alpha }
 \nonumber \\
&\times&
\left[\rho^{(\alpha -1)}(q_{N-1},N)-\rho^{(\alpha -1)}(q_{\alpha -1},N)e_{j,l}^* (q_{\alpha-1})\right]
\end{eqnarray}
Hence, the  density can be written as (see Eqs. (\ref{nonupart}) and (\ref{QGR1}))
\begin{eqnarray}
\label{eq1d}
 \rho (q_n)&=& \rho_0 +
\frac{1}{N^\alpha} 
\sum\limits_{\alpha =1}^N \sum_{j,l}^{[N]}
\frac{-e_{j,l}  (q_n) }{\left(1-e_{j,l}^*(q_1)\right)^\alpha } \nonumber\\
 &\times& 
\left[\rho^{(\alpha -1)}(q_{N-1},N)-\rho^{(\alpha -1)}(q_{\alpha -1},N)e_{j,l}^* (q_{\alpha-1})\right] \nonumber\\
 \end{eqnarray}

Let us now define the quantum Bernoulli polynomials
\begin{equation}
\label{BQdef}
B_{\alpha}(q_n,N) \equiv \frac{\alpha!}{N^\alpha} \sum_{j,l}^{[N]} \frac{-e_{j,l}  (q_n) }
{ \left(1-e_{j,l}^*(1/N)\right)^\alpha }
\end{equation}
(in Appendix \ref{app:P} we show that these are polynomials of degree $\alpha$).
With this definition, and re-writing the last term of Eq. (\ref{eq1d}) in terms of $B_{\alpha}(1-q_{n+1},N)$ we arrive to 
\begin{eqnarray}
\label{eq1d'}
 \rho (q_n)&=& \rho_0 \\
&+& \sum\limits_{\alpha =1}^N  \frac{1}{\alpha!} \left[B_{\alpha}(q_n,N) \rho^{(\alpha -1)}(q_{N-1},N)
 -   (-1)^\alpha  B_{\alpha}(1-q_{n+1},N) \rho^{(\alpha -1)}(q_{\alpha -1},N) \right] \nonumber
  \end{eqnarray}
This expression is the quantum version of the classical spectral representation  written in the form of Eq. (\ref{grepq}).  It is a quantum version of the Euler-Maclaurin expansion.  Note that, in contrast to the classical expansion, the quantum expansion is not symmetric with respect to the exchange of the points $x=0$ and $x=1$ (corresponding to $n=0$ and $n=N-1$). This is due to the difference operation we have used. It might be possible to use a symmetric difference operation, but this will not be investigated here.

%%%%%%%%%%%%%%%
\section{The quantum analogue of the generalized spectral representation for $\tau>0$}
\label{QG}
%%%%%%%%%%%%%%%%%

In this section we will obtain an expression describing the time evolution of density matrices produced by successive applications of the quantum Bernoulli map. We will obtain an expansion analogous to Eq. (\ref{grepqt}), which will involve the time-evolved quantum Bernoulli polynomials. Strictly speaking our expansion will not be a spectral representation of the quantum FP operator,  because the states we will construct are not eigenstates of the FP operator. However, these states will approach the classical eigenstates in the classical limit.

The time-evolution of the quantum Bernoulli polynomials is obtained from Eq. (\ref{UBQe}) as 
\begin{eqnarray}
\label{BQt}
(U_B^Q)^\tau B_{\alpha}(q_n,N) =  \frac{\alpha!}{N^\alpha} \sum_{j,l}^{[N]} \frac{1}{\mu_{j,l}(n,N)} \frac{-{\tilde e}_{j-\tau,l}  (q_n) }
{ \left(1-e_{j,l}^*(1/N)\right)^\alpha } \theta(j-\tau) 
\end{eqnarray}
Changing $j-\tau\to j$ and using Eq. (\ref{etild}) this becomes
\begin{eqnarray}
(U_B^Q)^\tau B_{\alpha}(q_n,N) = \frac{\alpha!}{N^\alpha} \sum_{j,l}^{[N/2^\tau]} \frac{\mu_{j,l}(n,N)}{\mu_{j+\tau,l}(n,N)} \frac{-{e}_{j,l}  (q_n) }
{ \left(1-e_{j+\tau,l}^*(1/N)\right)^\alpha } \quad {\rm for}\, 2^\tau \le N
\end{eqnarray}
and
\begin{eqnarray}
(U_B^Q)^\tau B_{\alpha}(q_n,N) = 0 \quad {\rm for}\, 2^\tau > N
\end{eqnarray}
Hereafter we consider the case $2^\tau \le N$. Using $\mu_{j+\tau,l}(n,N)=\mu_{j,l}(n,N/2^\tau)$ and $e^*_{j+\tau,l}(1/N)=e^*_{j,l}(2^\tau/N)$ we obtain
\begin{eqnarray}
\label{qfract}
(U_B^Q)^\tau B_{\alpha}(q_n,N) 
= \frac{1}{2^{\tau\alpha}}B_{\alpha}(q_n,N,\tau) 
\end{eqnarray}
where
\begin{eqnarray}
\label{qfract'}
B_{\alpha}(q_n,N,\tau) 
= -  \frac{\alpha!}{(N/2^\tau)^\alpha} \sum_{j,l}^{[N/2^\tau]}  
  \frac{\mu_{j-\tau,l}(n,N/2^\tau)}{\mu_{j,l}(n,N/2^\tau)} \frac{e_{j,l}  (q_{n}) }
{ \left(1-e_{j,l}^*(2^\tau/N\right)^\alpha } 
\end{eqnarray}
are the time-evolved quantum Bernoulli polynomials. In this expression we have (see Eq. (\ref{weight_s}))
\begin{equation}\label{qcorrodd}
  \frac{\mu_{j-\tau,l}(n,N/2^\tau)}{\mu_{j,l}(n,N/2^\tau)} 
  =
  \left\{ \begin{array}{ll}
        1 , & {\rm for\,} n \, {\rm even}\\
        1 + O(2^\tau k/N),  &  {\rm for\,} n \, {\rm odd}.
                 \end{array} \right.
 \end{equation}
with $k=2^j(2l+1)$. Thus for $n$ even the time-evolved Bernoulli polynomials in Eq. (\ref{qfract'}) have the same form as the initial polynomials in Eq. (\ref{BQdef}), except for the change $N\to N/2^\tau$. For $n$ odd there is an explicit correction coming from the ratio of weight factors in Eq. (\ref{qcorrodd}).

Let us introduce the re-scaled variables
\begin{equation}
   n' =  n/2^\tau, \quad N'=N/2^\tau
\end{equation}
then with $q_{n'}' \equiv n'/N' =  q_n = n/N$ we have
\begin{eqnarray}
\label{qfract2}
B_{\alpha} (q_n,N,\tau)&=& B_{\alpha} (q_{n'}', N'), \quad {\rm for\,} n {\rm \, even}\\
&=& B_{\alpha} (q_{n'}', N') + {\rm ``quantum" correction,} \quad {\rm for\,} n {\rm \, odd}\nonumber
\end{eqnarray}
where the ``quantum'' correction comes from the second line of Eq. (\ref{qcorrodd}).  This correction will play an important role in the next section when we will discuss the quasi-fractal shape developed by the time-evolved quantum Bernoulli polynomials.

Inserting Eq. (\ref{qfract}) into Eq. (\ref{eq1d'}) we get the following expression for the time evolution of the density matrix, expressed in terms of the time-evolved quantum Bernoulli polynomials: 
\begin{eqnarray}
\label{qgsr}
 (U_B^Q )^\tau \rho (q_n)&=& \rho_0 + \sum\limits_{\alpha =1}^N  \frac{1}{2^{\tau\alpha}}  \frac{1}{\alpha!} \left[B_{\alpha}(q_n,N,\tau) \rho^{(\alpha -1)}(q_{N-1}, N) \right. \nonumber\\
&-& \left.  (-1)^\alpha  B_{\alpha}(1-q_{n+1},N,\tau) \rho^{(\alpha -1)}(q_{\alpha -1},N) \right] 
  \end{eqnarray}
This equation is the quantum analogue of Eq. (\ref{grepqt}).
To see how the classical limit is reached, we write an alternative form of the quantum  Bernoulli polynomials
\begin{eqnarray}
\label{BQsin}
&&   B_{\alpha}(q_{n'}', N') = -\frac{\exp(\pi i n')}{(2N')^\alpha} \\
&+&  \sum_{k=1}^{N'/2} \frac{\exp\left[2\pi i k (n' +\alpha/2)/N'\right]
 + (-1)^\alpha \exp\left[-2\pi i k (n' +\alpha/2)/N'\right]}{\left[2iN' \sin(\pi k/N')\right]^\alpha} \nonumber
   \end{eqnarray}
 We may approximate
\begin{equation}
\label{sink}
2iN' \sin(\pi k/N') \approx 2\pi i k 
  \end{equation}
 provided that
\begin{equation}\label{ass1}
N' \gg 1.
 \end{equation}
In this case the Fourier components of the Bernoulli polynomials are dominated by  small  $k$ components with 
$ k \ll N'$.   Moreover let us assume that 
\begin{equation}\label{ass2}
\alpha \ll N'
 \end{equation}
Under these conditions the Fourier components for small $k$ are independent of $N'$ because $N'$ appears only in the expression  $n'/N'$, which is equal to $n/N$. This means that we can approximate
\begin{equation}
\label{qfract3}
B_{\alpha } (q_{n'}',N')  \approx  B_{\alpha} (q_n,N)
\end{equation}
and therefore, neglecting the correction in Eq. (\ref{qfract2}), we get
\begin{equation}
\label{qfract2'}
B_{\alpha} (q_n,N,\tau) \approx B_{\alpha } (q_n,N)
\end{equation}
Inserting this into Eq. (\ref{qfract}) we get
\begin{eqnarray}
\label{qfract_}
(U_B^Q)^\tau B_{\alpha}(q_n,N) 
\approx \frac{1}{2^{\tau\alpha}}B_{\alpha}(q_n,N) 
\end{eqnarray}
which corresponds to Eq. (\ref{UBB}). 
So the quantum Bernoulli polynomials behave as the classical ones, i.e., as decaying eigenstates of the FP operator.  Furthermore have (with $\alpha$ finite)
\begin{eqnarray}
\label{clim}
 \lim_{N'\to\infty} B_{\alpha}(q_{n'}', N') =   \lim_{N'\to\infty} \sum_{k=1}^{N'/2} \frac{\exp\left[2\pi i kn' /N'\right]
 + (-1)^\alpha \exp\left[-2\pi i k n'/N'\right]}{\left[2\pi i k \right]^\alpha}
   \end{eqnarray}
The right-hand side is an expression of the classical Bernoulli polynomial $B_\alpha(q_n)$. 

At $\tau=0$, provided only small values of $\alpha$ with $\alpha \ll N$ contribute to the summation in Eq.  (\ref{eq1d'}), the discrete difference goes to the continuous derivative  in the limit $N\to \infty$ (see Eq. (\ref{dev2diff})).  We recover the classical Euler-Maclaurin expansion (\ref{grepq}).  For $\tau>0$ and  $N\to\infty$ in Eq. (\ref{qgsr}) we recover the generalized spectral representation in Eq.  (\ref{grepqt}).

We remark that the existence of quantum decaying states  may be interpreted as a  signature of quantum chaos. The decomposition (\ref{qgsr})  shows that any density will approach equilibrium with the decay rates $1/2^{\alpha}$.  For $2^\tau = N$ the quantum Bernoulli polynomials  vanish identically and equilibrium is reached.

%%%
\section{Quasi-fractals}
\label{QF}
%%%

We consider now the quantum corrections in the Bernoulli map and the development of a quasi-fractals shape in the evolving quantum Bernoulli polynomials.
  
Let us first  now assume that $n$ (in $q_n$) is even such that $n' =  n/2^\tau$ is an integer. Keeping the assumptions (\ref{ass1}) and  (\ref{ass2}),  both $B_{\alpha}(q_n,N)$ and $B_{\alpha}(q_{n'}',N')$   give
a representation of the quantum Bernoulli polynomials with different number of grid points, namely $N$ and $N'$. The point $q_n=n/N=q_{n'}'=n'/N'$ belongs to both grids. If both $N, N' \gg 1$, then $B_{\alpha}(q_n,N) \approx B_{\alpha}(q_{n'}',N')$. After re-scaling by $2^{t \alpha }$, the quantum Bernoulli polynomials remain approximately constant at the points $q_n$ where  $n/2^\tau$ is integer.  Moreover, as discussed in the previous section they are approximately equal to their classical counterparts. They behave as decaying eigenstates of the FP operator.

In contrast, if $n$ is odd or more generally, if $n'=n/2^\tau$ is not an integer for $\tau>0$, then $q_{n'}'$ will not belong to the grid with $N'$ points. We expect a deviation from the  classical Bernoulli polynomial. This deviation is due to the discretization of space, so it will  give a quantum correction of the order of $\hbar$. In addition there will appear the quantum correction in Eq. (\ref{qfract2}).

Writing $n= 2^\gamma(2 r + 1)$ with both $\gamma>0$ and $r$ integers  the condition that  $n' = n/2^\tau$ is integer is translated as $\gamma \ge t$.  We denote the set of integers $n$ satisfying this condition as  $S_\tau$. We have
\begin{eqnarray}
\label{Sets}
S_0  &=& \{0,1,2,3,4,5,6,7,8, ...\} \nonumber\\
S_1  &=& \{0,2,4,6,8, ...\} \nonumber\\
S_2  &=& \{0,4,8, ...\}\nonumber\\
         &\cdots&
\end{eqnarray}
At each time step, the quantum Bernoulli polynomials act as  quasi-eigenstates of the FP operator only at the points in the sets $S_\tau$. At the other points  we get deviations.   The  recursive nature of the sets in Eq. (\ref{Sets}) means that these deviations appear in a  self-similar fashion. This gives the evolving Bernoulli polynomials a quasi-fractal shape. 

\begin{figure}
\begin{center}
\includegraphics[width=4in, angle=270]{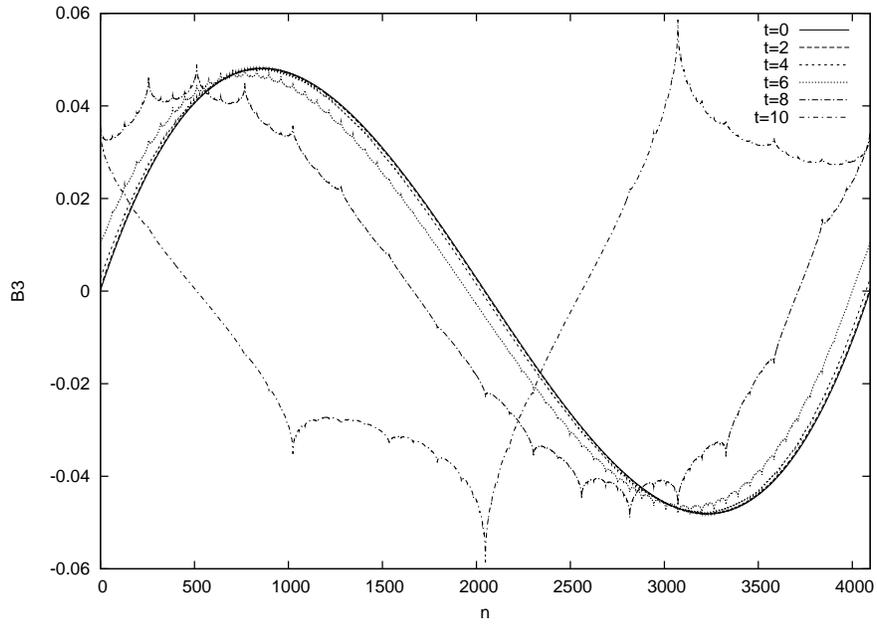} 
\end{center}
\caption{
Plot of $(2^3U_B^Q )^\tau B_{3} (q_n,N )$ for $N = 4096$
}
\label{fig3}
\end{figure}

In order to visualize this, we have employed a computer program to calculate the exact evolution of densities under the quantum Bernoulli map. As an example, at $\tau=0$ we take as the initial state the polynomial $B_{\alpha} (q_n,N)$ with $\alpha=3$ and $N=4096$.  At each step, we re-scale the density by $2^3$. If this were the classical Bernoulli map, the graphs would remain unchanged for $\tau>0$, because $B_3(x)$ is an eigenstate of the FP operator with eigenvalue $1/2^3$.  But for the quantum map we have a different behavior. The result is shown in Figure \ref{fig3}.
%%%
%%%
%%%
\begin{figure}
\begin{center}
\includegraphics[width=3in, angle=270]{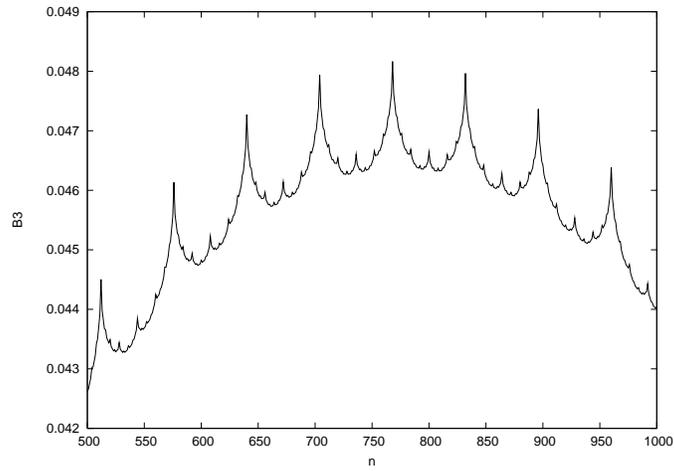} 
\end{center}
\caption{
Plot of $(2^3U_B^Q )^\tau B_{3} (q_n,N )$  for $N = 4096$ and $\tau = 6$
}
\label{fig4}
\end{figure}
%%%
\begin{figure}
\begin{center}
\includegraphics[width=3in, angle=270]{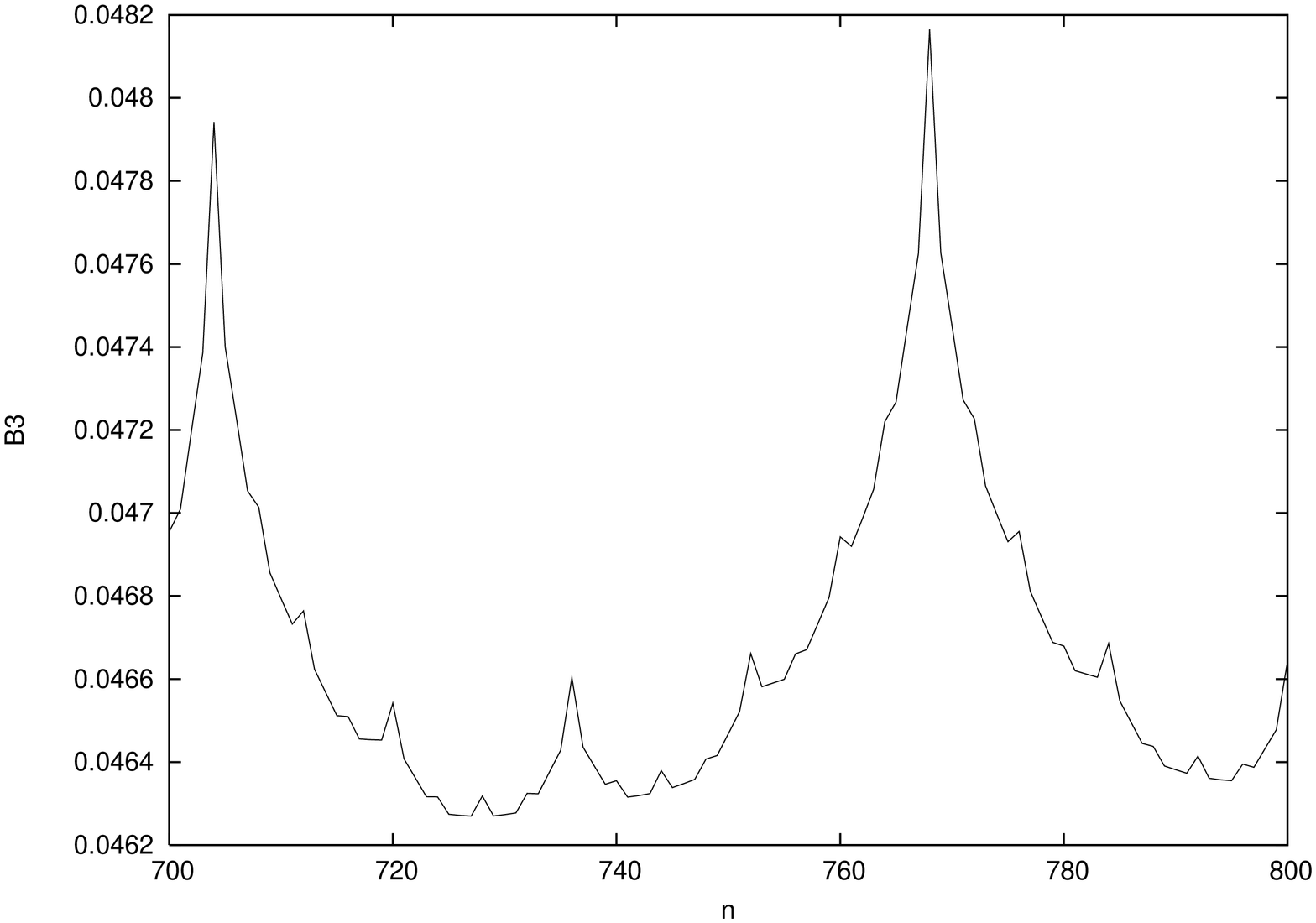} 
\end{center}
\caption{
Zoom-in of Figure \ref{fig4}
}
\label{fig5}
\end{figure}
%%%
We see the graphs developing  the quasi-fractal  shape mentioned above.  We also notice that as $\tau$ increases the graphs are shifted. This is due to the $\alpha$ term in the numerator of  the right hand side of Eq. (\ref{BQsin}). Provided $\alpha \ll N' =  N/2^\tau$ the shift is negligible, but as $\tau$ increases the shift becomes noticeable. 

\begin{figure}
\begin{center}
\includegraphics[width=3in, angle=270]{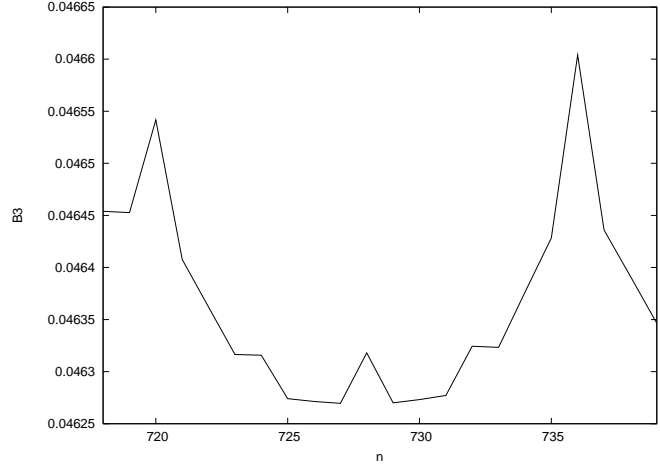} 
\end{center}
\caption{
Zoom-in of Fig. \ref{fig5}. If we zoom-in this figure, 
the self-similarity disappears.}
\label{fig6}
\end{figure}

Figure \ref{fig4} shows a zoom-in view of $B_{\alpha} (q_n,N)$ with $\alpha=3$, $N=4096$ and $\tau=6$. Figures \ref{fig5}-\ref{fig6} show further zoom-in views.
%%%
%%%

The self-similarity of the graphs stops when the space resolution is of the order of $1/N \sim \hbar$ (see  figure \ref{fig6}). For this reason, we call this evolving state a quasi-fractal.

\begin{figure}
\begin{center}
\includegraphics[width=3in, angle=270]{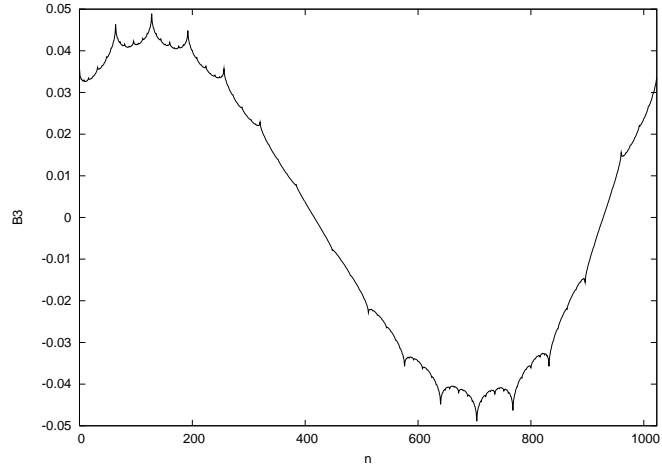} 
\end{center}
\caption{
Plot of $(2^3U_B^Q )^\tau B_{3} (q_n,N )$  for $N =1024$ and $\tau =6$}
\label{fig8}
\end{figure}

\begin{figure}
\begin{center}
\includegraphics[width=3in, angle=270]{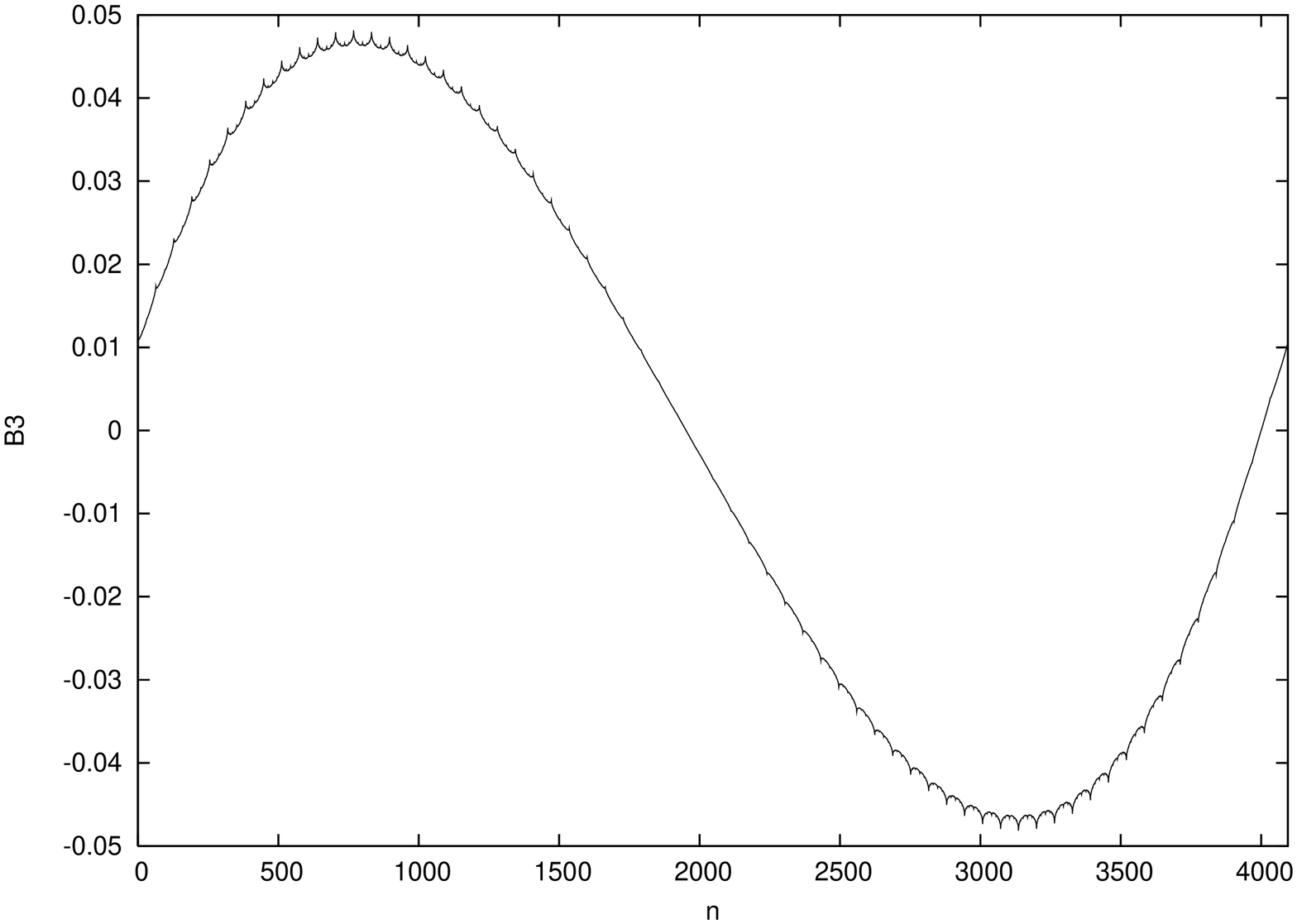} 
\end{center}
\caption{
Plot of $(2^3U_B^Q )^\tau B_{3} (q_n,N)$ for $N =4096$ and $\tau =6$ (compare with Fig. \ref{fig8}). As $N$ increases ($\hbar$ decreases), the curve looks smoother. At the same time, the self-similar pattern appears at smaler length scales.}
\label{fig7}
\end{figure}

When $N\to\infty$, the quasi-fractal behaves as a true fractal.
 At the same time, when $N\to\infty$, the amplitude of the fractal deviations from the classical Bernoulli polynomial gets smaller and the graph looks smoother. Eventually, it just looks like the classical Bernoulli polynomial (see
Figures \ref{fig8} and  \ref{fig7}). 
%%%
%%%
%%%

%%%%%%%%%%%%%%
\section{Concluding remarks }
\label{CR}
%%%%%%%%%%%%%%

We summarize the main results. We introduced the quantum Bernoulli map by coupling the quantum baker map to an external environment  that produces instant decoherence after each time step.

We described the quantum Bernoulli map  in terms of decaying quasi-eigenstates of the Frobenius-Perron  operator. These states are analogous to the classical Bernoulli polynomials.  We found  conditions under which  these quantum states approach the classical ones.  Moreover we found a quantum analogue of the generalized spectral representation of the classical Bernoulli map. We also found a quantum analogue of the Euler-Maclaurin expansion. In this expansion the classical Bernoulli polynomials are replaced by their quantum counterparts; derivatives are replaced by differences. 
 
We have suggested that a signature of quantum chaos may be the presence of decaying eigenstates (or quasi-eigenstates) like the ones we have described in this paper.

One interesting finding is that even after decoherence, the quantum Bernoulli map shows quantum corrections with respect to the classical Bernoulli map.  The quantum corrections give a quasi-fractal shape to the evolving quantum Bernoulli polynomials.   These corrections vanish in the classical limit. At the same time, as the classical limit is approached, the quasi-fractal  approaches a true fractal. In a sense, this fractal  is hidden in the classical limit. To our knowledge this is a  feature chaotic quantum systems that  has not received  attention before.

An open question is:  do quasi-fractals like the ones we described here appear in other chaotic quantum systems? Self-similarity in quantum systems has been discussed in other contexts. For example in \cite{Reichl,Reichl2}, self-similarity has been discussed in the context of non-linear resonances.  Refs.  \cite{Tasaki6,Tasaki7,Tasaki8} have discussed singular spectra or fractal spectra of quantum systems, that appear, for example, with quasi-periodic lattices. Ref. \cite{Berry} gives an interesting discussion on fractals generated by a quantum rotor. The self-similarity discussed in the present work is of a different origin. Still, it would be worthwhile to investigate any connections with these other works.

Another question for future work is whether the quantum baker map can be described along the lines developed in this paper (i.e. using quantum Bernoulli polynomials and difference operators).

A final point worth noting is that the model we have described may be realized as a set of shifting quantum bits subject to repeated measurements after each shift. The measurements  cause instantaneous decoherence.  If the measurement time interval is greater than the Zeno time \cite{Misra}, the system decays exponentially \cite{Yingyue2}. Equations (\ref{qfract})  and (\ref{qfract_}) manifest this property. However if  the time interval between measurements approaches zero,  we might obtain the quantum Zeno effect. We speculate that repeated measurements will  prevent  the system from decaying despite the underlying chaotic dynamics.

\bigskip
 {\bf Acknowledgements}  {We thank Hiroshi Hasegawa,  Dean Driebe, Linda Reichl, Tomio  Petrosky and George Sudarshan for helpful  suggestions and discussions. G. O. will always be grateful for Prof. Tasaki's friendliness, hospitality and generosity throughout the years, and for having learned so many interesting physics ideas from him.}

\appendix
%%%%%%%%%%%%%%
\section{The quantum Bernoulli polynomials}
\label{app:P}
%%%%%%%%%%%%%%
In this appendix we show that the functions
\begin{equation}
\label{BQdefA}
B_{\alpha,N}(q_n) = - \frac{\alpha!}{N^\alpha} \sum_{j,l}^{[N]} \frac{-e_{j,l}  (q_n) }
{ \left(1-e_{j,l}^*(q_1)\right)^\alpha }
\end{equation}
are polynomials of degree $\alpha$ in $q_n$. Let us first state two properties of these functions, which are easily proved:
\begin{equation}
\label{Bpol1}
\sum_{n=0}^{N-1} B_{\alpha,N}(q_n) = 0
\end{equation}
and (for $n>0$)
\begin{eqnarray}
\label{Bpol2}
 B_{\alpha,N}(q_n) -  B_{\alpha,N}(q_{n-1}) &=& (\alpha/N) B_{\alpha-1,N}(q_n), \quad \alpha>1\nonumber\\
 B_{1,N}(q_n) -  B_{1,N}(q_{n-1}) &=& 1/N, \quad n>0
\end{eqnarray}
From the last equation we find that
\begin{equation}
\label{Bpol3}
  B_{1,N}(q_n) = q_n + B_{1,N}(q_0)
\end{equation}
We can find $B_{1,N}(q_0) $ using  Eq. (\ref{Bpol3}) together with Eq.  (\ref{Bpol1}), which gives  
\begin{equation}
\label{Bpol4}
 \sum_{n=1}^N  q_n + N B_{1,N}(q_0)  = 0.
\end{equation}
Using
\begin{equation}
\label{Bpol5}
 \sum_{n=1}^{N-1} n = \frac{N^2}{2} - \frac{N}{2}
\end{equation}
we get
\begin{equation}
\label{Bpol6}
  B_{1,N}(q_n) = q_n - \frac{1}{2} + \frac{1}{2N}
\end{equation}
In a similar way, using
\begin{equation}
\label{Bpol7}
 \sum_{n=1}^{N-1} n^2 =  \frac{N^3}{3} - \frac{N^2}{2} + \frac{N}{6}
\end{equation}
we find that
\begin{equation}
\label{Bpol8}
  B_{2,N}(q_n) = q_n^2 - q_n + \frac{1}{6} + \frac{2}{N} q_n - \frac{1}{N} + \frac{5}{6N^2} 
\end{equation}
We can continue in this way for $\alpha=3, 4, \cdots$ showing that $ B_{\alpha,N}(q_n)$ are polynomials of degree $\alpha$. Moreover, 
recalling that $1/N \sim \hbar$ and consulting a table of classical Bernoulli polynomials \cite{Dean}, we find that
\begin{equation}
\label{Bpol9}
  B_{\alpha,N}(q_n) = B_\alpha(q_n) + O(\hbar)
\end{equation}
for $\alpha=1,2$. Similar relations must hold for $\alpha>2$ with $\alpha\ll N$, because of Eq. (\ref{clim}).

%%%%%%%%%%%%%%%%%%%%
% References       %
%%%%%%%%%%%%%%%%%%%%

\end{document}